\documentclass[preprint,showpacs,preprintnumbers,amsmath,amssymb,nofootinbib]{revtex4}
\usepackage{epsfig}
\usepackage{dcolumn}
\usepackage{bm}

\begin{document}

\title{Proton-Neutron Coupling in the Gamow Shell Model: the Lithium Chain}

\author{N. Michel$^{1-3}$}
  \email{michel@mail.phy.ornl.gov}
\author{W.~Nazarewicz$^{1,2,4}$}
  \email{witek@utk.edu}
\author{M. P{\l}oszajczak$^{5}$}
   \email{ploszajczak@ganil.fr}

\affiliation{%
$^1$Department of Physics and Astronomy, University of Tennessee,
Knoxville, TN 37996 \\
$^2$Physics Division, Oak Ridge National Laboratory, P.O.~Box 2008,
Oak Ridge, TN 37831 \\
$^3$Joint Institute for Heavy-Ion Research, Oak Ridge, TN 37831 \\
$^4$Institute of Theoretical Physics, Warsaw University, ul.~Ho\.{z}a
69, 00-681 Warsaw, Poland \\
$^5$Grand Acc\'{e}l\'{e}rateur National d'Ions Lourds (GANIL),CEA/DSM -- 
CNRS/IN2P3, BP 55027, F-14076 Caen Cedex 05, France
}
\date{\today}

\begin{abstract}
The shell model in the complex $k$-plane (the so-called Gamow Shell Model)
has recently been formulated and applied to 
structure of weakly bound, neutron-rich nuclei. The completeness relations of Newton and Berggren,
which apply to the neutron case, are strictly valid  for  finite-range potentials. 
However, for long-range potentials, such as the Coulomb potential for protons,
for which the arguments based on  the  Mittag-Leffler theory do not hold, the completeness
still needs to be  demonstrated. This has been done in this paper, both analytically and
numerically. 
The generalized Berggren relations are  then used in the first Gamow Shell Model study 
of nuclei having {\it both}  valence neutrons and protons, namely the 
 lithium chain. The single-particle basis used is that of the Hartree-Fock-inspired
potential generated  by a finite-range residual interaction. The effect of isospin mixing
in excited unbound states is discussed.
\end{abstract}


\maketitle

\section {Introduction}

One of the main frontiers of the nuclear many-body problem is the structure
of exotic, short-lived nuclei with extreme neutron-to-proton ratios.
Apart from intrinsic nuclear structure interest, properties of these
nuclei are crucial for our understanding of astrophysical processes
responsible for cooking of elements in stars.
 From a
theoretical point of view, the major challenge
 is to achieve a consistent picture of structure and reaction aspects 
of weakly bound and  unbound nuclei, which requires an accurate description of
the particle continuum \cite{(Dob98)}. Here, the tool of choice is the
continuum shell model (see Ref.~\cite{(Oko03)} for a recent review) and, most recently, the 
Gamow Shell Model (GSM) \cite{(Mic02),(Mic03),(Mic04),(Mic04a)} 
(see also Refs.~\cite{(Bet02),(Bet03),(Bet04)}). 
GSM is the multi-configurational shell model with a single-particle (s.p.)
basis given by the Berggren ensemble \cite{(Ber68),(Ber93),(Lin93)} which consists of
 Gamow (or
resonant) states and the complex non-resonant continuum. The resonant states
are the generalized eigenstates of the time-independent Schr\"{o}dinger
equation which are regular at the origin and satisfy purely outgoing boundary
conditions. The s.p. Berggren basis is  generated by a 
finite-depth potential, and the many-body states are obtained in
shell-model  calculations as the linear combination of Slater determinants
spanned by resonant and non-resonant s.p. basis states. Hence, both
continuum effects and correlations between nucleons are taken into 
account simultaneously. The interested reader can find all details of the
formalism in Ref. \cite{(Mic03)}, in which
the GSM was applied to many-neutron configurations in neutron-rich
helium and oxygen isotopes.

When extending the GSM formalism to the general neutron-proton case, with 
both protons and neutrons occupying valence s.p. states, 
one is confronted with a theoretical problem: the Berggren
completeness relations, which are the pillars of the Gamow Shell Model,
have been strictly proved (and checked numerically
\cite{(Lio96),(Mic03)}) only for quickly vanishing (finite-range)
local potentials, while the repulsive 
 Coulomb potential for protons has infinite range.
 The theoretical problem lies, in fact, not in the Berggren
(complex-energy) completeness relation itself, but in the Newton (real-energy)
completeness
relation \cite{(New82),(New02)}. This latter involves both 
bound and scattering states, upon
which the Berggren completeness relations can be demonstrated
using the method of analytic continuation.

Our paper is organized as follows.
Section~\ref{completo} contains a derivation of the Newton completeness
relation that is valid for a rather wide class of  potentials, including
the Coulomb potential. Based on this result,
the Berggren
completeness relation for protons is derived in the same way as previously done for
  neutrons \cite{(Mic03)}. The numerical tests of
the completeness  of the proton Berggren ensemble is given in Sec.~\ref{WSWS}. 
Section \ref{GHF} introduces the  Hartree-Fock (HF) inspired procedure
used to optimize  the s.p. basis, the so-called Gamow-HF method.
The   first GSM calculation involving  active neutrons and protons 
 is presented in Sec.~\ref{lithiums},
with the the $1p$-shell study of the lithium chain, ranging from  $^5$Li to
$^{11}$Li. The residual interaction used is a surface-peaked
finite range force. A novel aspect,
absent in our previous GSM studies, is the appearance
of $T$=0 couplings which seem to exhibit significant particle-number 
(or density) dependence.
Finally, Sec.~\ref{concludo} contains  the main conclusions of our work.

\section{Completeness Relations in the GSM: Analytical Considerations}\label{completo}

As the one-body completeness relation  for resonant and scattering states 
 is prerequisite for our theory, we shall
 demonstrate it rigorously. We shall first
consider the case of a local potential and then generalize it to 
nonlocal potentials.

\subsection{Local potential}

In order to demonstrate the orthonormality and completeness relations for 
s.p. proton states, we consider a spherical proton potential that is
finite at $r$=0, and it has  a pure Coulomb behavior for $r \rightarrow
+\infty$. The one-body radial wave functions  $u(r)$
are solutions of  the Schr\"odinger
equation:
\begin{eqnarray}
& & u''(r) = \left[ \frac{l(l+1)}{r^2} + v(r) - k^2 \right] u(r) \label{local_Schrodinger_equation} \\
& & v(r) \sim \frac{\mbox{const}}{r} \mbox{ , } r \rightarrow +\infty \label{Coulomb_asymp}
\end{eqnarray}
where potential $v$ is given in units of  fm$^{-2}$, and $l$ is the angular momentum of the particle.
Let us consider the bounded region enclosed in a large sphere of radius $R$. (This can be
achieved by introducing  an infinite  well of radius $R$ surrounding the nucleus.)
Of course, in the final result, $R$ will be allowed to go to infinity.
For each value of $R$, one has the following completeness relation on $[0:R]$ \cite{(Mor53)}: 
\begin{eqnarray}\label{boxset}
\sum_{n \in b} u_{n}(r) u_{n}(r') + 
\sum_{m=0}^{+\infty} u^{(d)}(k_m,r) u^{(d)}(k_m,r') = \delta (r - r'),
\end{eqnarray}
where $b$ denotes the set of bound states having radial
wave functions $u_n(r)$ with $ k_n^2 < v(R)$, and
 $u^{(d)}(k_m,r)$ is a wave function of a
 normalized discretized continuum state, given by the 
boundary  conditions $u^{(d)}(k_m,0) = 0$ and $u^{(d)}(k_m,R) = 0$.

For the purpose of this discussion, 
it is convenient to introduce the set of wave functions \cite{(Lin93)}
\begin{equation}\label{redef_scat}
u(k_i,r) = \frac {u^{(d)}(k_i,r)}{\sqrt{k_{i+1} - k_i}},
\end{equation}
which obey the following normalization condition:
\begin{equation}\label{ndelt}
\langle u(k_i) | u(k_j) \rangle = \frac {\delta_{ij}}{k_{i+1} - k_i}.
\end {equation}
Since, in addition,
\begin{eqnarray}
& & \langle u(k) | u_n \rangle = 0 \\
& & \displaystyle \langle u_n | u_{n'} \rangle = \delta_{nn'},
\end{eqnarray}
the
box completeness relation  can be written as: 
\begin{eqnarray}
\sum_{n \in b} u_{n}(r) u_{n}(r') + \sum_{m=0}^{+\infty} u(k_m,r) u(k_m,r') (k_{m+1} 
- k_m) = \delta (r - r'). \label{box_compl}
\end{eqnarray}
When $R \rightarrow +\infty$, 
the infinite series  in Eq.~(\ref{box_compl})   becomes an integral, thus 
giving the expected completeness relation.
Unfortunately, this cannot be done right away, as the series 
and the integral converge only in a weak sense. 

To prove the convergence rigorously, 
let us  consider the  completeness relation of the free box expressed
in the form of Eq.~(\ref{redef_scat}):
\begin{equation}
\sum_{m=0}^{+\infty} B_m^2  \hat{j}_l(\kappa_m r) \hat{j}_l(\kappa_m r') 
(\kappa_{m+1} - \kappa_m) = \delta (r - r'), \label{free_box_compl}
\end{equation}
where $\hat{j}_l(\kappa_m r')$ are the
 Riccati-Bessel functions,  $\hat{j}_l(\kappa_m R)$=0 ($m$=0,1,2...),
and  $B_m$ is the normalization constant:
\begin{equation} \label{free_norm}
- B_m^2
\frac{R}{2} \hat{j}_{l+1}(\kappa_m R)
\hat{j}_{l-1}(\kappa_m R)  = \frac {1}{\kappa_{m+1} - \kappa_m}.
\end{equation}
Subtracting (\ref{free_box_compl}) from (\ref{box_compl}), one obtains: 
\begin{eqnarray}
& & \sum_{n \in b} u_{n}(r) u_{n}(r') \nonumber \\ 
&+& \sum_{m=0}^{+\infty} \left[ u(k_m,r) u(k_m,r') 
 (k_{m+1} - k_m) - B_m^2  \hat{j}_l(\kappa_m r) \hat{j}_l(\kappa_m r') (\kappa_{m+1} - \kappa_m) \right] 
\nonumber \\
&=& 0. \label{box_cv_series}
\end{eqnarray}
We shall now demonstrate  that the above series converges
 in the sense of functions, so the limiting transition  from a series to an 
 integral when $R \rightarrow +\infty$ can be easily carried out.  

To this end, let us consider  the behavior of the $m$-th term 
in  the series when $m$ (and $k_m$) $\rightarrow +\infty$.
For very large values of $k_m$, one can use the semiclassical expansion  
in powers of $k_m^{-1}$ : 
\begin{eqnarray}
u(k_m,r) &=& C_m \hat{j}_l(k_m r) - C_m \frac{\mathcal{V}(r)}{2 k_m} \hat{j}_l'(k_m r) + O 
\left( \frac{C_m}{k_m^2} \right), \label{WKB_Bessel} \\
\hat{j}_l(k_m r) &=& \sin \left( k_m r - l \frac{\pi}{2} \right) 
- \frac{a_l}{2 k_m r} \cos \left( k_m r - l \frac{\pi}{2} \right) + O \left( \frac{1}{k_m^2 r^2} \right), \label{Bessel_exp}
\end{eqnarray}
where $C_m$ is a normalization constant, $a_l$ is a constant depending on $l$ only,
and
\begin{equation}\label{lshift}
 \mathcal{V}(r) =\int_{0}^{r} v(r') \; dr'.
\end{equation}
For the Coulomb potential, $\mathcal{V}(r) \propto \ln{r}$; hence the expression
(\ref{WKB_Bessel}) properly accounts for the logarithmic term in the phase shift.
It immediately follows from Eqs.~(\ref{WKB_Bessel}) and (\ref{Bessel_exp}) that
\begin{eqnarray}
k_m &=& \frac{ \left( m + \frac{l}{2} \right) \pi}{R} + \frac{a_l + R \mathcal{V}(R)} {2 R m \pi} + O \left( \frac{1}{m^2} \right), \label{k}  \\
\kappa_m &=& \frac{\left( m + \frac{l}{2} \right) \pi}{R} + \frac{a_l} {2 R m \pi} + O \left( \frac{1}{m^2} \right). \label{free_k}
\end{eqnarray}
The constant
$C_m$ can be  determined from the normalization condition:
\begin{equation}
 C_m^2 \int_{0}^{R} \left[ \hat{j}^2_l(k_m r)- 
\frac{\mathcal{V}(r)}{k_m} \hat{j}_l(k_m r) \hat{j}_l'(k_m r) \right] dr  
+ O \left( \frac{C_m^2}{k_m^2} \right)
=  \frac {1}{k_{m+1} - k_m}. \label{norm}
\end{equation}
Since the  integral
 involving $\mathcal{V}$  behaves
 like $1/k_m^2$, 
 $C_m$  becomes:
\begin{equation}\label{jl_const}
C_m^2  \frac{R}{2} \left[ \hat{j}^2_l(k_m R) -
\hat{j}_{l+1}(k_m R) \hat{j}_{l-1}(k_m R) \right] + O \left(
\frac{C_m^2}{k_m^2} \right) = \frac {1}{k_{m+1} - k_m},
\end{equation}
cf. Eq.~(\ref{free_norm}).

Using Eqs.~(\ref{k}) and (\ref{free_k}), one  obtains:
\begin{eqnarray}
C_m &=& \sqrt{\frac{2}{\pi}} + O \left( \frac{1}{m^2} \right), \label{norm_result} \\
B_m &=& \sqrt{\frac{2}{\pi}} + O \left( \frac{1}{m^2} \right). \label{free_norm_result} 
\end{eqnarray}
Let us consider the behavior of $B_m$ for $R \rightarrow
+\infty$ but $\kappa_m \rightarrow \kappa$ with $\kappa$$>$0. The
expansion of Eq.~(\ref{Bessel_exp}) is still valid in this case, as for
$r$=$R$, it is a familiar expansion in $1/R$ of the Bessel function. It 
follows from Eq.~(\ref{Bessel_exp}) that:
\begin{equation} 
\kappa_m  = \frac{\left( m + \frac{l}{2} \right)
\pi}{R} + O \left( \frac{1}{R^2} \right) \label{free_k_R_big}
\end{equation} 
with $m$ chosen so $\kappa_m$ is the closest to $\kappa$.
Then, Eqs.~(\ref{free_norm}) and (\ref{Bessel_exp}) give:
\begin{equation} 
B_m = \sqrt{\frac{2}{\pi}} + O \left( \frac{1}{R} \right)
\label{free_norm_result_big_R},
\end{equation}
as expected.

Note that the leading term in Eqs.~(\ref{norm_result}) and (\ref{free_norm_result})
is the familiar  normalization of continuum wave functions 
\cite{(Baz69)}. The reminders, of the order of $m^{-2}$,
guarantee the convergence of the series.
By using Eqs.~(\ref{WKB_Bessel})-(\ref{free_k}),
one can show that the series (\ref{box_cv_series}) converges
 for all $r > 0$ and $r' > 0$.

As a consequence, in the limit of  $R \rightarrow +\infty$, Eq.~(\ref{box_cv_series})
becomes:
\begin{eqnarray}
\sum_{n \in b} u_{n}(r) u_{n}(r') 
+ \int_{0}^{+\infty} \left[ u(k,r) u(k,r') - {\frac{2}{\pi}} \hat{j}_l(k r) \hat{j}_l(k r') \right] \; dk = 0. \label{cv_integral}
\end{eqnarray}
By taking advantage of the closure relation for the Riccati-Bessel functions,
\begin{eqnarray}
\int_{0}^{+\infty}  \hat{j}_l(k r) \hat{j}_l(k r') \; dk = {\frac{\pi}{2}} \delta(r - r') \label{real_Bessel_completeness},
\end{eqnarray}
one finally arrives at the sought completeness relation:
\begin{eqnarray}
\sum_{n \in b} u_{n}(r) u_{n}(r') + \int_{0}^{\infty} u(k,r) u(k,r') \; dk = \delta(r - r'). \label{real_completeness_relation}
\end{eqnarray}
By using the same arguments as in  Ref.~\cite{(Mic03)}, one obtains
the generalized Berggren completeness relation, also valid for the proton case:
\begin{eqnarray}
\sum_{n \in b,d} u_{n}(r) u_{n}(r') + \int_{L^{+}} u(k,r) u(k,r') \; dk = 
\delta(r - r'). \label{complex_completeness_relation}
\end{eqnarray}
For details, including 
the numerical treatment of scattering wave functions and
corresponding  matrix elements, we refer the reader to  Ref.~\cite{(Mic03)}.
Let us only  remark, in passing, that in the presence of the Coulomb potential
the standard regularization procedure \cite{(Zel60),(Zel61)} has to be
modified \cite{(Dol77)}. In our work, however, we apply the 
exterior complex scaling method \cite{(Gya71),(Sim79)} which works very well
regardless of whether the Coulomb potential is used or not.

\subsection{Nonlocal potential}\label{nnloc}
In the presence  of a nonlocal potential, such as the HF  exchange  potential 
generated by a 
finite-range two-body interaction, the Schr\"odinger equation
 (\ref{local_Schrodinger_equation}) becomes:
\begin{eqnarray}
u''(r) &=& \left[ \frac{l(l+1)}{r^2} + v_l(r) - k^2 \right] u(r) + \int_{0}^{+\infty} v_{nl}(r,r') u(r') \; dr' \label{non_local_Schrodinger_equation}
\end{eqnarray}
where $v_l$ is the local part of the potential, and $v_{nl}$ its nonlocal kernel. 
We assume that  $v_{nl}(r,r') \rightarrow 0$ when $r \rightarrow +\infty$ or 
$r' \rightarrow +\infty$ (nuclear potential has to be localized) and that
$v_{nl}(r,0) = 0$ $\forall$ $r$ (the  potential is regular at the origin). As the radial
HF functions are regular at zero, the latter
condition is automatically met  for the
HF exchange potential.

If the  integral containing the nonlocal potential $v_{nl}$ behaves like $1/k^2$ when
$k \rightarrow +\infty$, then 
the asymptotic expression 
(\ref{WKB_Bessel})  holds. Indeed, integration by  parts yields:
\begin{eqnarray}
& & \int_{0}^{+\infty} v_{nl}(r,r') \left[ j_l(kr') - \frac{\mathcal{V}_l(r') j_l'(kr')}{2k} \right] \; dr'  \nonumber\\
&=& \frac{1}{k^2} \left[ \frac{\partial v_{nl}}{\partial r'} (r,0)  \mathcal{J}_l(0) 
 + \int_{0}^{+\infty} \frac{\partial^2 v_{nl}}{\partial r'^2} (r,r')  \mathcal{J}_l(k r') \; dr' \right] + O \left( \frac{1}{k^2} \right) \\
&=& O \left( \frac{1}{k^2} \right), \nonumber
\end{eqnarray}
where  $\displaystyle \mathcal{V}_l(r) = \int_{0}^{r} v_l(r') \; dr'$, 
$\displaystyle \mathcal{J}_l(t) = \int_{t_0}^{t} \int_{t_0^{'}}^{t'} j_l(t'') \; dt'' \; dt'$, and 
$t_0$ and $t'_0$ are chosen so $\mathcal{J}_l(t)$ is bounded on $[0:+\infty[$.
Consequently, Eq.~(\ref{WKB_Bessel}) also  holds for  nonlocal potentials.
The proof of completeness can be, therefore,  performed in the same way as for local potentials,
by simply replacing  $v$ by $v_l$ in all expansions in $k^{-1}$.

\section{Completeness of the one-body proton Berggren ensemble: numerical tests} \label{WSWS}

In this section, we shall discuss examples of the Berggren completeness relation
in the one-proton case (for  the neutron case, see Ref.~\cite{(Mic03)}). The s.p. basis is
generated by the spherical
Woods-Saxon (WS) -- plus -- Coulomb potential:
\begin{eqnarray}
V(r) &=& -V_0 f(r) - 4 V_{\rm so} {\bm l} \cdot {\bm s} \frac{1}{r} \frac{df(r)}{dr} + V_c(r), \\
f(r) &=& \left[ 1 + \exp \left( \frac{r-R_0}{d} \right) \right]^{-1}.
\end{eqnarray}
In all the examples of this section, the WS potential has the radius $R_0$=5.3 fm,
diffuseness $d$=0.65 fm,
and the spin-orbit strength $V_{\rm so}$=5~MeV.
The Coulomb potential $V_c$ is assumed to be generated by  a uniformly charged sphere of radius $R_0$ 
and charge $Q$=+20$e$.
The depth of the central part is  varied to simulate different situations. 

In this section, we
shall expand the    $2p_{3/2}$ state, $|u_{\rm WS}\rangle$,
either weakly bound or resonant, in
the basis $|u_{\rm WS^B}(k)\rangle$ generated by the WS potential of a different depth:
\begin{eqnarray}\label{expanded_state_continous}
|u_{\rm WS}\rangle &=& \sum_{i}c_{k_i}|u_{\rm WS^B}(k_i)\rangle \nonumber \\
 &+& \int_{L_+}  c(k) |u_{\rm WS^B}(k)\rangle\,dk,
\end{eqnarray}
cf. Eq.~(\ref{complex_completeness_relation}).
In the above equation, the first term in the expansion
represents contributions from the resonant states while the second term 
is the non-resonant continuum contribution. 
Since the basis is properly normalized, the expansion amplitudes meet the
condition:
\begin{equation}\label{cexpnorm}
\sum_{i}c^2_{k_i} + \int_{L_+}  c^2(k)\,dk = 1.
\end{equation}
In all cases considered, 
the $0p_{3/2}$ and $1p_{3/2}$ orbitals  are well bound  
(by $\sim 50$ MeV and $\sim 20$ MeV, respectively) and do not play any 
significant role in the
 expansion studied,  although they are
taken into account in the actual calculation. The $2p_{3/2}$ state is, however,
either loosely bound or resonant, and  the scattering states 
along the contour  ${L_+}$  are essential to guarantee  the completeness.
 To take the non-resonant continuum into account,  we take
the complex contour ${L_+}$ that corresponds to three straight segments in
the complex $k-$plane, joining the points: $k_0 = 0.0-i0.0$, 
$k_1=0.3-i0.1$, $k_2 = 1.0-i0.0$, and $k_3 = 2.0-i0.0$ (all in fm$^{-1}$). The contour is
discretized with $n$=40 points:
\begin{equation}
|2p_{3/2} \rangle \simeq \sum_{n \in b,d} c_n |u_{n} \rangle + \sum_{i=1}^{n} c_{k_i} |u(k_i) \rangle.  \label{expanded_state_discrete}
\end{equation}

\begin{figure}[htbp]
\begin{center}
\includegraphics[width=10cm]{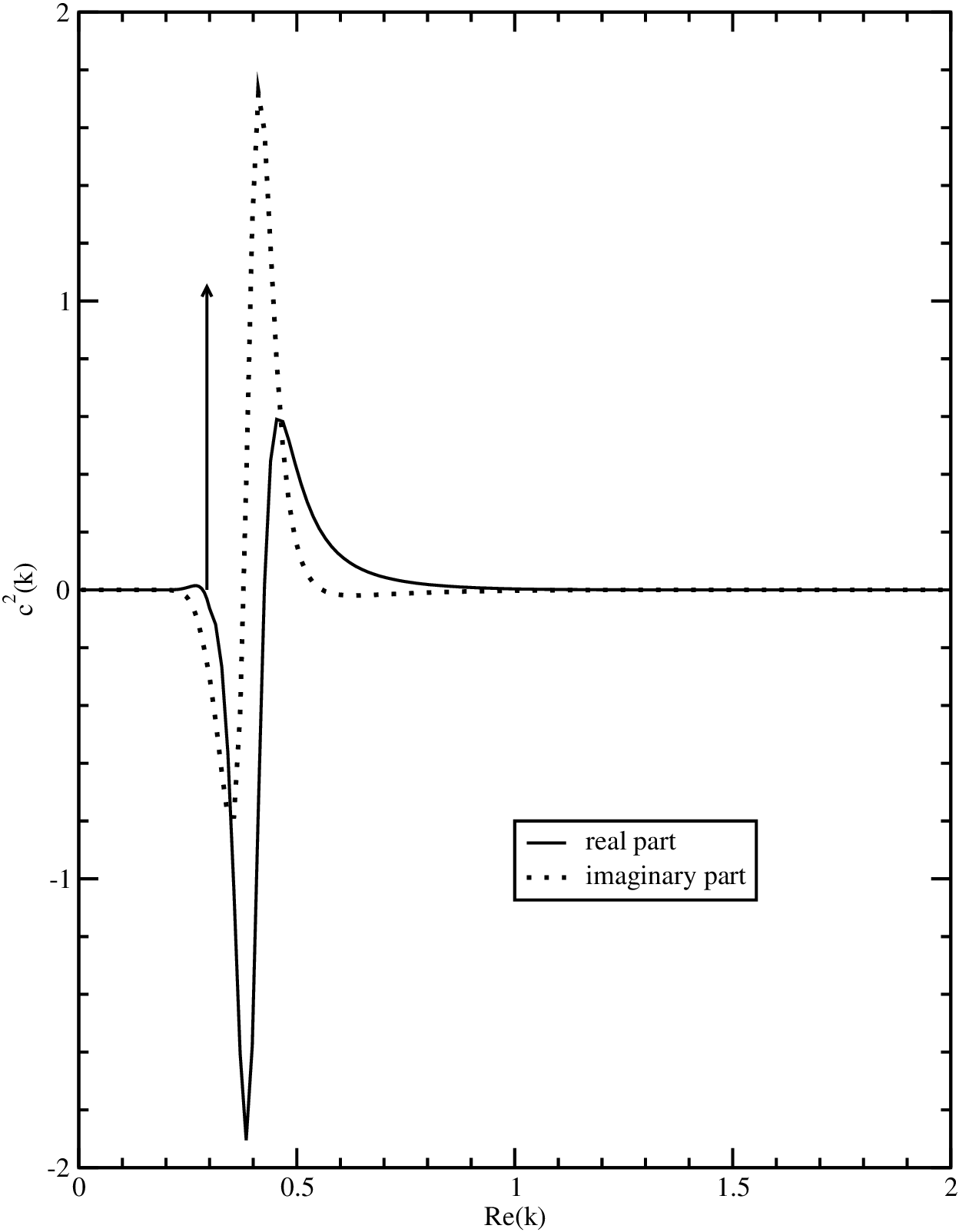}
\end{center}
\caption{Distribution of the squared amplitudes $c^2(k)$ of the
$2p_{3/2}$ proton state of a Woods-Saxon  potential with a depth $V_0=65$ MeV, in the
s.p. basis generated by a Woods-Saxon potential with a depth $V_0=70$ MeV. 
The Coulomb potential  is assumed to be that of a uniformly charged sphere.
The amplitudes of both real (solid line) and imaginary (dotted line) components
of the wave function are plotted as a function of $\Re[k]$. The height of
the arrow gives the squared amplitude of the $2p_{3/2}$ state contained
in the basis.}
\label{coeff_70_65}
\end{figure}

In the first example, we shall expand the $2p_{3/2}$ s.p. resonance ($E=$3.287 MeV, $\Gamma=$931 keV)
of a WS potential of the depth $V_0$=65 MeV in the basis generated by the WS 
potential of the depth  $V_0^{B}$=70 MeV. (Here the $2p_{3/2}$ s.p. resonance has
an energy $E=$1.905 MeV and width $\Gamma=$61.89 keV.) 
After diagonalization in the discretized basis  (\ref{expanded_state_discrete}),
one obtains
$E=$3.289 MeV and $\Gamma=$934 keV for the $2p_{3/2}$ s.p. resonance, i.e.,
the discretization error is $\sim$3 keV.
The density of the expansion 
amplitudes is shown in Fig.~\ref{coeff_70_65}.
As both states are resonant, the squared amplitude  of the
$2p_{3/2}$ basis state is close to one. Nevertheless, the contribution
from the non-resonant
continuum is essential. It is due to
the fact that the resonant state in the basis is very narrow, whereas the expanded
 resonant state is fairly broad. It is interesting to
notice that the contribution from scattering states with energies smaller than that
of the resonant state is practically negligible; 
this  is due to the confining effect of
the Coulomb barrier. 

\begin{figure}[htbp]
\begin{center}
\includegraphics[width=10cm]{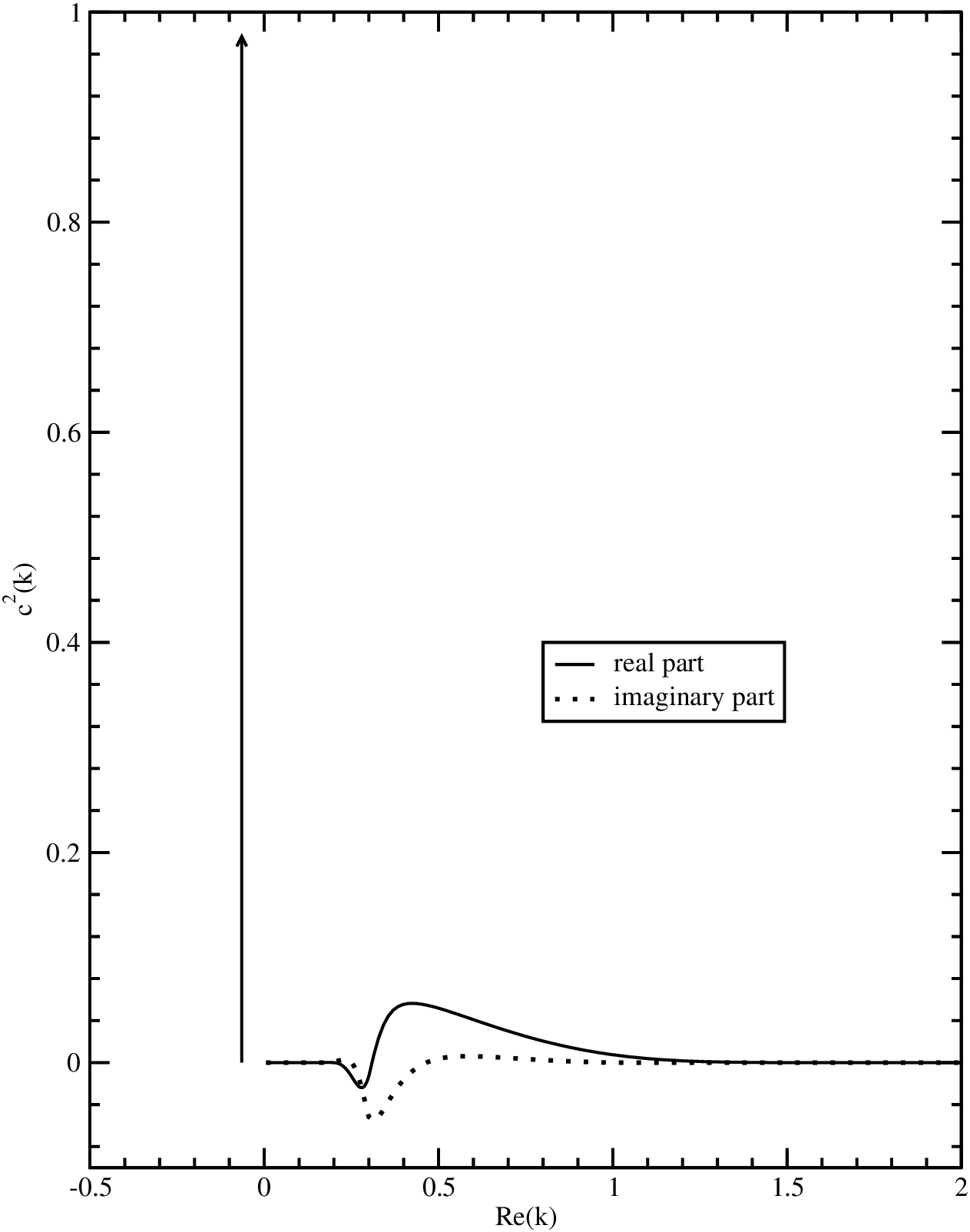}
\end{center}
\caption{Similar as in Fig. \ref{coeff_70_65} except for the bound
$2p_{3/2}$ s.p. state of the  WS potential with $V_0$=75 MeV
expanded in the  basis generated by another WS
potential ($V_0^{(B)}$ = 80 MeV). The height of the arrow gives 
the squared amplitude  of the bound $2p_{3/2}$ state at the  value
of --$\Im[k]$. (The corresponding 
$k-$value is purely imaginary.)}
\label{coeff_75_80}
\end{figure}

The second example, shown in Fig.~\ref{coeff_75_80},
deals with the case of a  $2p_{3/2}$ state that
 is bound in both potentials. Here  $V_0$=75 MeV and 
$V_0^{(B)}$=80 MeV, and the  $2p_{3/2}$  state lies at $E$=--0.0923  MeV  and $E$=--2.569 MeV,
respectively.
Here, the scattering component  is almost negligible, which reflects the localized character 
of bound proton states. After  the diagonalization,
one obtains $E$=--2.568 MeV and $\Gamma$=1.73 keV for the $2p_{3/2}$ state,
 which is indeed very close to the exact  result. 

In the third example, the unbound 
$2p_{3/2}$ state ($E=$1.905 MeV and $\Gamma=$61.89 keV;
$V_0$=70 MeV) is expanded in a WS basis containing
the bound $2p_{3/2}$ level ($E$=--0.0923 MeV; $V_0^{(B)}$=75 MeV).
 As a consequence, the resonance's width has to be brought  by the
 scattering states. Nevertheless, the component of the $2p_{3/2}$
state of the basis is still close to one, whereas the continuum component 
plays a secondary role. Once again,  one
can see a Coulomb barrier
effect: even if the expanded $2p_{3/2}$ state is unbound, its wave
function is very localized due to the large Coulomb barrier; hence it
has a large overlap with the bound $2p_{3/2}$ basis state.  The diagonalization yields
$E=$1.9065 MeV and $\Gamma=$58.9 keV, i.e., the discretization error
is again close to 3\,keV.
\begin{figure}[htbp]
\begin{center}
\includegraphics[width=10cm]{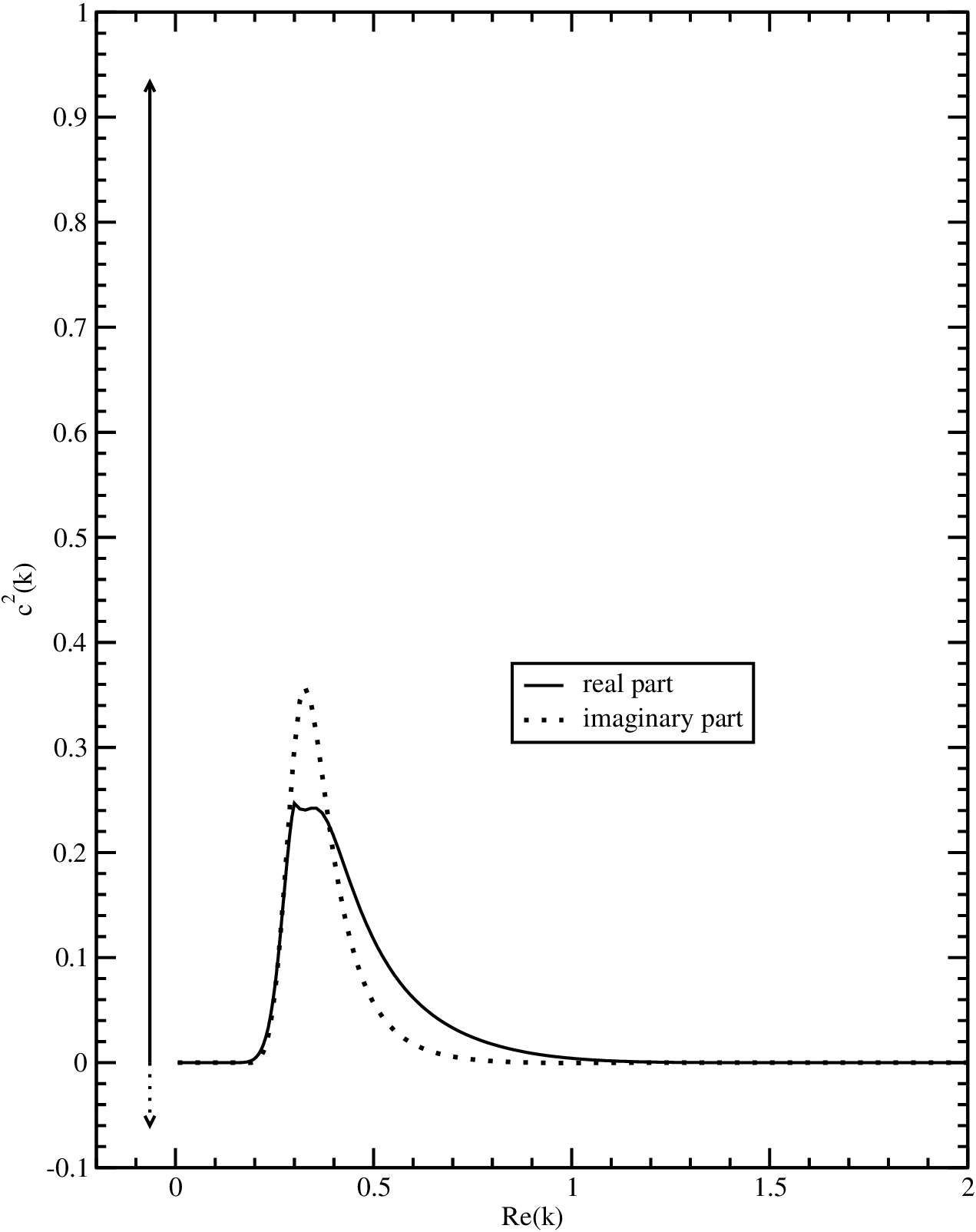}
\end{center}
\caption{Similar as in Fig. \ref{coeff_75_80}
except for the 
$2p_{3/2}$ resonance of the  WS potential with $V_0$=70 MeV
expanded in the  basis generated by another WS
potential ($V_0^{(B)}$ = 75 MeV).}
\label{coeff_75_70}
\end{figure}

\begin{figure}[htbp]
\begin{center}
\includegraphics[width=10cm]{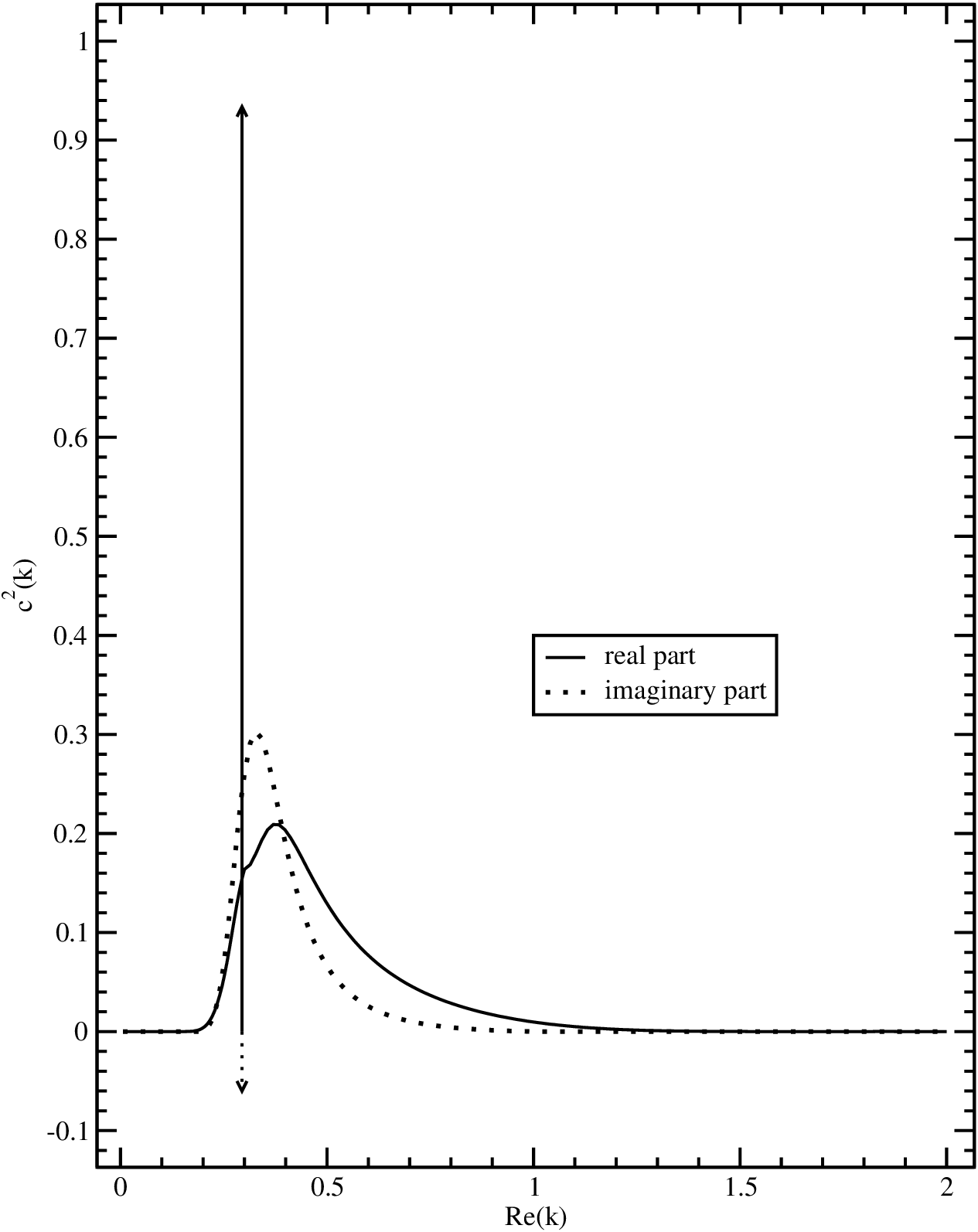}
\end{center}
\caption{Similar as in Fig. \ref{coeff_75_80}
except for the 
$2p_{3/2}$ resonance of the  WS potential with $V_0$=75 MeV
expanded in the  basis generated by another WS
potential ($V_0^{(B)}$ = 70 MeV).} 
\label{coeff_70_75}
\end{figure}

In the last example, shown in Fig.~\ref{coeff_70_75},
 the basis is that of
the WS potential with a depth of 70 MeV, and the expanded state corresponds to a
WS potential  with $V_0$=75 MeV. 
This is the most interesting  case since one expresses
a bound (real) state in the basis which contains only complex wave
functions. (The contribution from well bound $0p_{3/2}$ and $1p_{3/2}$ s.p. 
states is negligible.)
As expected, the behavior of the amplitudes 
is very close to that of the previous example
 (see Fig. \ref{coeff_75_70}). Moreover, one can notice  that the
scattering states become important in the expansion when their energies
approach  the resonant state energy. The diagonalization gives 
$E=$--0.0925 MeV
and $\Gamma=$-3.8 keV for the $2p_{3/2}$ state.

Summarizing this section, our numerical tests demonstrate  that the 
one-body Berggren completeness relation works very well in the
 proton case involving the Coulomb potential. For other numerical tests,
 see Refs.~\cite{(Ver98)} (study of the s.p. level density) and
 \cite{(Lin94)}   (study of the Berggren expansion in the pole
 approximation).

\section{Spherical Gamow Hartree-Fock method}\label{GHF}

In our previous calculations of the He chain 
\cite{(Mic02),(Mic03)}, we used the s.p. basis of the WS potential representing
 $^5$He. This basis is appropriate at the beginning of the He chain, but when 
   departing from the core nucleus, its quality
deteriorates. For instance, the ``$^5$He" basis is not expected to be optimal 
 when applied to the neutron-rich  halo nucleus $^{8}$He because of
 the very different asymptotic behavior of this weakly bound system.
 The obvious remedy is to use the s.p.
 basis that is optimal for a given nucleus, that is the HF basis.
However, since the Berggren ensemble used in GSM is required to possess spherical symmetry,
HF calculations must be constrained to spherical shapes. Moreover, since
in some cases one is interested in unbound nuclei lying beyond the drip line
(particle resonances), the HF procedure has to be extended to 
unbound  states. In the following, the HF-based procedure
that meets the above criteria is referred to as
the  Gamow-Hartree-Fock (GHF) method.

Since, strictly speaking,  the spherical HF potential  cannot be defined 
for open shell nuclei, one has to resort to approximations.
In this work, we tried two different ways of averaging the HF potential. The first
ansatz is  the usual uniform-filling approximation in which 
 HF occupations are averaged over individual spherical shells. In the second ansatz,
the deformed HF potential corresponding to non-zero angular
momentum projection
is averaged over all the magnetic quantum numbers.
 Both methods reduce to the
true HF potential in the case of closed-shell nuclei.

\subsection{Average spherical HF potential}\label{HFpot}

In  the uniform-filling approximation
 no individual
 HF orbitals are blocked. 
 The matrix elements of the HF potential
 $U_{\rm uf}$ between two spherical states $\alpha$
and $\beta$ carrying quantum numbers ($j,l$) are:
\begin{eqnarray}
\langle \alpha | U_{\rm uf} | \beta \rangle 
&=& \langle \alpha | \hat{h} | \beta \rangle \nonumber \\
  &+& \frac{1}{2 j+1} \sum_{m,\lambda,m_{\lambda}} \frac{N(\lambda)}{2j_{\lambda}+1}
    \langle \alpha m \; \lambda m_{\lambda} | \hat{V} | \beta m \; \lambda m_{\lambda} \rangle
\end{eqnarray}
where  $\hat{h}$
is the s.p.  Hamiltonian (given by a WS+Coulomb potential),  $\lambda$ is an
occupied shell with  angular quantum numbers ($j_{\lambda},l_{\lambda}$), 
$N(\lambda)$ is the number of nucleons occupying this shell,
and $\hat{V}$ is the residual shell-model interaction.

To define the $M$-potential, one occupies (blocks) 
the s.p.  states in the valence shell that have the largest 
 angular momentum projections on the third axis. The resulting
 Slater determinant corresponds to the  angular momentum $J$=$M$. 
For closed-shell nuclei ($M$=0)
and  for  nuclei with one particle
 (or hole) outside a closed subshell ($M$=$j$), this Slater determinant can be associated
with the ground state of the s.p. Hamiltonian. However, in other cases
it corresponds to   an excited state with $J$$>$0.
Spherical
$M$-potential, $U_M$, is defined by  averaging the resulting HF potential
 over  magnetic quantum number $m$:
\begin{eqnarray}
\langle \alpha | U_M | \beta \rangle 
&=& \langle \alpha | \hat{h} | \beta \rangle  \nonumber \\
  &+& \frac{1}{N_{l,j}} \sum_{m=j+1-N_{l,j}}^{j} \sum_{\lambda}
   \langle \alpha m \;\lambda m_{\lambda} | \hat{V} | \beta m \; \lambda m_{\lambda} \rangle
\end{eqnarray}
where 
$N_{l,j}$ is the number of nucleons occupying the valence shell
with  quantum numbers $l,j$. 

The $M$-potential is expected to work better for nuclei with one particle
(or hole)  outside the
closed subshell.
However, one can expect this potential to be
not as good as  $U_{\rm uf}$  when the Slater determinant with $J$=$M$  
represents an excited state.

\subsection{Unbound HF states}

While the  HF procedure 
 is well defined for the bound states, it has to
be modified for the unbound s.p. states (resonant or scattering),
 even in the case of closed-shell
nuclei. First, the effective
nuclear two-body interaction has to be  quickly vanishing
beyond  certain radius. Indeed, if it does not, the resulting
HF potential diverges when $r
\rightarrow +\infty$, thus providing incorrect s.p. asymptotics. 
Moreover, as resonant states are complex, the true self-consistent  HF
potential is complex. This is to be avoided, as the Berggren
completeness relation assumes a real potential. However,
since we are interested
in the optimal basis-generating potential and  not in the full-fledged complex-energy
HF problem \cite{(Kru97)},  we simply take  the real part of the
 (generally complex) HF potential.

\subsection{Treatment of the exchange part} 

As the residual interaction used in our shell-model calculations 
is finite range, its exchange part gives rise to a non-local
potential. The HF equations solved in the
coordinate space can be written as
integro-differential equations. The standard method to
treat such a HF  problem is by means of the equivalent local potential
\cite{(Vau57)}:
\begin{equation}\label{eq_pot}
V_{\rm eq}(r) = v_{l}(r) + \frac{\int_{0}^{+\infty} v_{nl}(r,r') u(r') \; dr'} {u(r)}, 
\end{equation}
where we use the notation of Eq.~(\ref{non_local_Schrodinger_equation}).
The resulting HF equations are local but 
potentials become  state dependent.  The main difficulty is the appearance
of singularities in $V_{\rm eq}(r)$  due to the zeroes of the s.p. wave function. 
This problem is practically solved
by replacing $u(r)$ with a small number (e.g.,  $10^{-3}$) in the
denominator of Eq.~(\ref{eq_pot}) when  $u(r)$ approaches  zero, and by using
splines to define the HF potential. The numerical accuracy is
checked by calculating   overlaps between different wave functions
having the same ($j,l$) values. The overlaps are  typically $10^{-5}$,
which is small enough to consider  wave functions orthogonal.

Let us note in passing that 
the  demonstration of Sec.~\ref{nnloc} is valid when
 the non-local part of the potential is localized,
as it is the case for the nuclear interaction.
However, if one wants to  explicitly consider
the Coulomb HF potential between valence protons, 
one has to resort to approximations in order
to avoid its infinite-range non-local part. One possibility is
to use the so-called Slater approximation, which has been  shown
to work fairly well \cite{(Ska01)}. Another method, the so-called
generalized local approximation, 
has been proposed in Ref.~\cite{(Bul95)}, where the
 Coulomb exchange term has been parametrized
in terms of a coordinate-dependent effective mass.
In any  case, the effect of the approximate  treatment of the 
Coulomb exchange term on
 the GHF basis is very small  as compared to 
 other uncertainties related to  the construction of the  GHF Hamiltonian.
 
\subsection{Choice of the average potential}

In order to compare the quality of two GHF potentials, 
we inspect the binding energy (i.e., 
the expectation value of the GSM Hamiltonian) for several Li and He isotopes 
in the truncated GSM space.  As discussed in Sec.~\ref{lithiums} below,
we took at most two particles in the 
GHF continuum. Due to this truncation, and as well as the discretization
of the contour $L_+$ and the assumption of the momentum cut-off
(the contour does not extend to infinity),
the  completeness relation of the  resulting many-body
shell-model basis is violated and, except for some special cases, the results obtained
with different average potentials are different.

According to the variational principle, 
better  basis-generating potentials must  yield lower  binding energies. 
Table~\ref{BE_tables_Li} shows the  binding energies of  $^{6,7,9}$He
and $^{7-11}$Li calculated
in the GSM. Since  $^8$He and $^{10}$He are 
closed-shell nuclei,  both $U_{\rm uf}$ and $U_M$ are identical  in these cases.
\begin{table}
\caption{Binding energies of the He and Li isotopes (in MeV) calculated in the GSM
 using the GHF basis  with  (i) the  uniform-filling approximation potential  $U_{\rm uf}$
         or (ii)  the $U_M$-potential. See   Sec.~\protect\ref{HFpot} for definitions.}
\vskip 0.5truecm 
\label{BE_tables_Li}
\parbox{14cm}{    
\begin{ruledtabular}
\begin{tabular}{|c|cccccccc|} 
              & $^6$He & $^7$He  & $^9$He  & $^7$Li   & $^8$Li   & $^9$Li   & $^{10}$Li & $^{11}$Li \\ \hline
$U_{\rm uf}$         & -1.038 & -0.048 & -2.357  & -14.266  & -15.521  & -20.288  & -18.082   & -15.649   \\ 
$U_M$        & -0.984 & -0.475  & -2.418 & -13.008  & -15.094  & -20.181  & -17.749   & -15.634   
\end{tabular}
\end{ruledtabular}
}
\end{table}
One can see that the uniform-filling approximation for the GHF potential works 
better  for $^7$Li, $^8$Li, and $^{10}$Li, whereas the s.p. basis
of the  $M$-potential
is a better choice for $^7$He. In  all the remaining cases, the two potentials
give results that are practically equivalent. In the particular case of
  $^6$He and $^6$Li (not displayed), there are  {\em  at most} two
nucleons in the non-resonant continuum. Consequently, the shell-model
spaces for $^6$He and $^6$Li are almost complete,  and the results are
almost independent on the choice of  s.p. basis.
Based on our tests, in our GHF calculations we use  the uniform-filling approximation 
except for nuclei with one particle (hole) outside  
closed subshells where the $M$-potential provides a slightly more
optimal  s.p. basis.

\section{Gamow Shell Model description of the Lithium chain}\label{lithiums}

\subsection {Description of the calculation}
In our previous studies \cite{(Mic02),(Mic03)}, 
we have used the s.p. basis generated
by a WS potential which was adjusted to reproduce 
the s.p. energies in $^5$He. This  potential (``$^{5}$He" parameter set
\cite{(Mic03)}) is characterized by  the radius $R$=2
fm, the diffuseness $d$=0.65 fm, the strength of the central field 
$V_0$=47 MeV, and
the spin-orbit strength $V_{\rm so}$=7.5 MeV. 
As a residual interaction, we took 
the Surface Delta Interaction (SDI).
However, when it comes to practical applications, 
SDI  has several disadvantages.
Firstly, it has zero range, so 
an energy cutoff has to be introduced; hence the residual interaction
depends explicitly
on the 
model space.
Moreover, as the SDI interaction cannot practically 
be used to generate the  
HF potential (it produces a nonrealistic mean field), 
one is bound to use the same WS basis for all nuclei of interest,
which is far from optimal
as the number of valence particles increases.  
So, we have decided to introduce  \cite{(Mic04a)} a finite-range
residual interaction, the 
Surface Gaussian Interaction (SGI): 
\begin{equation}
V^{\rm SGI}_{J,T}({\bm r}_1, {\bm r}_2) = V_0(J,T) \cdot \exp \left[ - \left( \frac{{\bm r}_1-{\bm r}_2}{\mu} 
\right) ^2 \right]
\cdot \delta{ \left( |{\bm r}_1| +  |{\bm r}_2| - 2 \cdot R_0 \right) },
\label{eq1}
\end{equation}   
which is used together with the WS potential
with the ``$^{5}$He" parameter set.

The Hamiltonian employed in our work can thus be written as follows:
\begin{equation}\label{hamm}
\hat{H} = \hat{H}^{(1)} + \hat{H}^{(2)}
\end{equation}
where $\hat{H}^{(1)}$ is  the one-body Hamiltonian described above augmented by
a hard sphere Coulomb potential of radius $R_0$ from the $^4$He core, and $\hat{H}^{(2)}$ is the 
two-body  interaction
among valence particles, which can be written as a sum  of SGI and Coulomb terms.
It is important to emphasize that the  
Coulomb interaction between valence protons can be treated
as precisely at the shell-model level as the nuclear part. Indeed,
the Coulomb two-body matrix
elements  in the HF basis can be calculated using
the exterior complex scaling as
decribed in Ref.~\cite{(Mic03)}. 
Though not used in this paper, as we are  considering only one valence
proton in the shell model space, this feature of the Gamow Shell Model
allows for a precise treatment of the Coulomb term.

The SGI interaction
is a compromise between  the SDI and the Gaussian 
interaction. The parameter
$R_0$ in Eq.~(\ref{eq1}) is the radius of the WS potential, and 
$V_0(J,T)$ is the coupling constant which explicitly depends
on the total angular momentum $J$ and  the total isospin 
$T$ of the nucleonic pair.
A principal  advantage of the SGI is that it 
is finite-range, so no  energy cutoff is, in principle, needed.
Moreover, the surface delta term in (\ref{eq1}) simplifies 
the calculation of two-body matrix elements,
because they can be reduced to one-dimensional radial integrals.
 (In the  case of other finite-range interactions, such as  the Gogny force \cite{(Dec80)},  
 the radial integrals are two-dimensional.)
Consequently, with SGI, an adjustment of the Hamiltonian parameters
becomes feasible.  

The Hamiltonian (\ref{hamm}) is diagonalized in the Berggren basis
generated by means of the  GHF procedure of Sec.~\ref{GHF}.
This allows  one to use the optimal 
spherical GHF potential for each  nucleus studied; hence 
a more efficient
truncation in the space of configurations with a different 
number of particles in the non-resonant continuum.

\subsection {Choice of the valence space}

The valence space for protons and neutrons consists of 
the $0p_{3/2}$  and $0p_{1/2}$ GHF resonant states, calculated for  each nucleus,
and the  $\{ ip_{3/2}\} $ and $\{ ip_{1/2}\} $ $(i=1,\cdots ,n)$, respectively complex and real continua
generated by the same potential. These continua extend from $\Re[k]$=0 to
$\Re[k]$=8\,fm$^{-1}$, and they are  discretized with 14 points 
(i.e., $n$=14). The $0p_{1/2}$ state is taken into account only if it is bound or very narrow.
For the lightest isotopes considered, 
it is a very broad resonant state  ($\Gamma \sim$ 5 MeV),
and, on physics grounds,  it is more justified to simply take a real $\{ ip_{1/2}\} $ contour,
so the completeness relation is still fulfilled. 

Altogether, we  have
15 $p_{3/2}$ and 14 or 15 $p_{1/2}$  GHF  shells 
in the GSM calculation. 
The imaginary parts of $k$-values of the discretized 
continua are chosen to minimize the error
made in calculating the imaginary parts of energies of the many-body states.
Another continua, such as  $s_{1/2}$, $d_{5/2}$, $\cdots$ are neglected, 
as they can be chosen to be real and would only induce
a renormalization of the two-body interaction. We have checked  
\cite{(Mic02),(Mic03)} that their influence on the binding energy of light 
helium isotopes is negligible. On the other hand, the $1s_{1/2}$ anti-bound 
neutron s.p. state is important in the heaviest Li isotopes ($^{10}$Li, $^{11}$Li)
and plays a significant role in explaining the halo ground-state
(g.s.) configuration of 
$^{11}$Li \cite{(Tho94),(Bet04)}. At present, however, solving a GSM problem 
for $^{11}$Li  in the full $psd$ 
GHF space
is not possible within a reasonable computing time. 

Having defined a discretized GHF  basis,  we construct
the many-body Slater determinants from all s.p. basis states 
(resonant and scattering), 
keeping only those with at most two particles in the non-resonant continuum.
 Indeed, according to our tests, as the two-body Hamiltonian is
diagonalized in its optimal GHF basis, the weight of configurations involving
 more 
than two particles in the continuum is usually quite small,
and they are neglected in the following. To illustrate this point,
let us  consider the $J^\pi$=$3/2^{-}$ ground state
 of $^7$Li. Since, in this case, there are only three valence particle, 
  the complete calculation
is possible. As seen in Table~\ref{tables_GS_Li7_trunc},
the weight of the component  with three particles in the
non-resonant continuum, $L_{+}^{(3)}[\pi \nu^2]$ is indeed  two orders of magnitude smaller than other
configurations weights. The presence of $L_{+}^{(3)}[\pi \nu^2]$ 
modifies  the weights of other  configurations on a very ninor way. For
the leading
configuration, $0p_{3/2}[\pi]$$0p^{2}_{3/2}[\nu]$,
without any particles in the non-resonant continuum, the effect is $\sim$8\%. 
As a consequence, the neglect of the $L_{+}^{(3)}[\pi \nu^2]$ component  leaves
 the overall structure of the state
unchanged, and one can safely  truncate the shell model space while
slightly renormalizing the two-body residual interaction.
\begin{table}
\caption{\label{tables_GS_Li7_trunc}Squared amplitudes of different configurations in the ground 
state of $^7$Li without truncation (second column)
and with truncation to at most two particles in the 
non-resonant continuum (third column). The sum of squared amplitudes of all Slater determinants with 
$n$ particles (protons $[\pi]$ or/and neutrons $[\nu]$)
in the non-resonant continuum is denoted by $L_{+}^{(n)}$.}
\begin{center}
\parbox{13cm}{
\begin{ruledtabular}
\begin{tabular}{|c|c|c|} 
Configuration & {No Truncation}  & {Truncation} \\ \hline
$0p_{3/2}[\pi]$ $0p^{2}_{3/2}[\nu]$ & {0.561--i2.783$\cdot 10^{-4}$} &  {0.612--i2.285$\cdot 10^{-4}$}\\ 
$L_{+}^{(1)}[\pi]$   & {0.096+i4.732$\cdot 10^{-5}$} & {0.096+i4.980$\cdot 10^{-5}$}\\ 
$L_{+}^{(1)}[\nu]$   & {0.184+i1.203$\cdot 10^{-4}$} & {0.164+i1.077$\cdot 10^{-4}$} \\ 
$L_{+}^{(2)}[\nu^2]$   & {0.064+i2.419$\cdot 10^{-5}$} & {0.054+i1.600$\cdot 10^{-5}$} \\ 
$L_{+}^{(2)}[\pi \nu]$   & {0.088+i7.032$\cdot 10^{-5}$} & {0.075+i5.508$\cdot 10^{-5}$}\\ 
$L_{+}^{(3)}[\pi \nu^2]$   & {0.008+i1.621$\cdot 10^{-5}$}  & 0 
\end{tabular}
\end{ruledtabular}
}
\end{center}
\end{table} 


\newpage

\begin{figure}[htbp]
\begin{center}
\includegraphics[width=10cm]{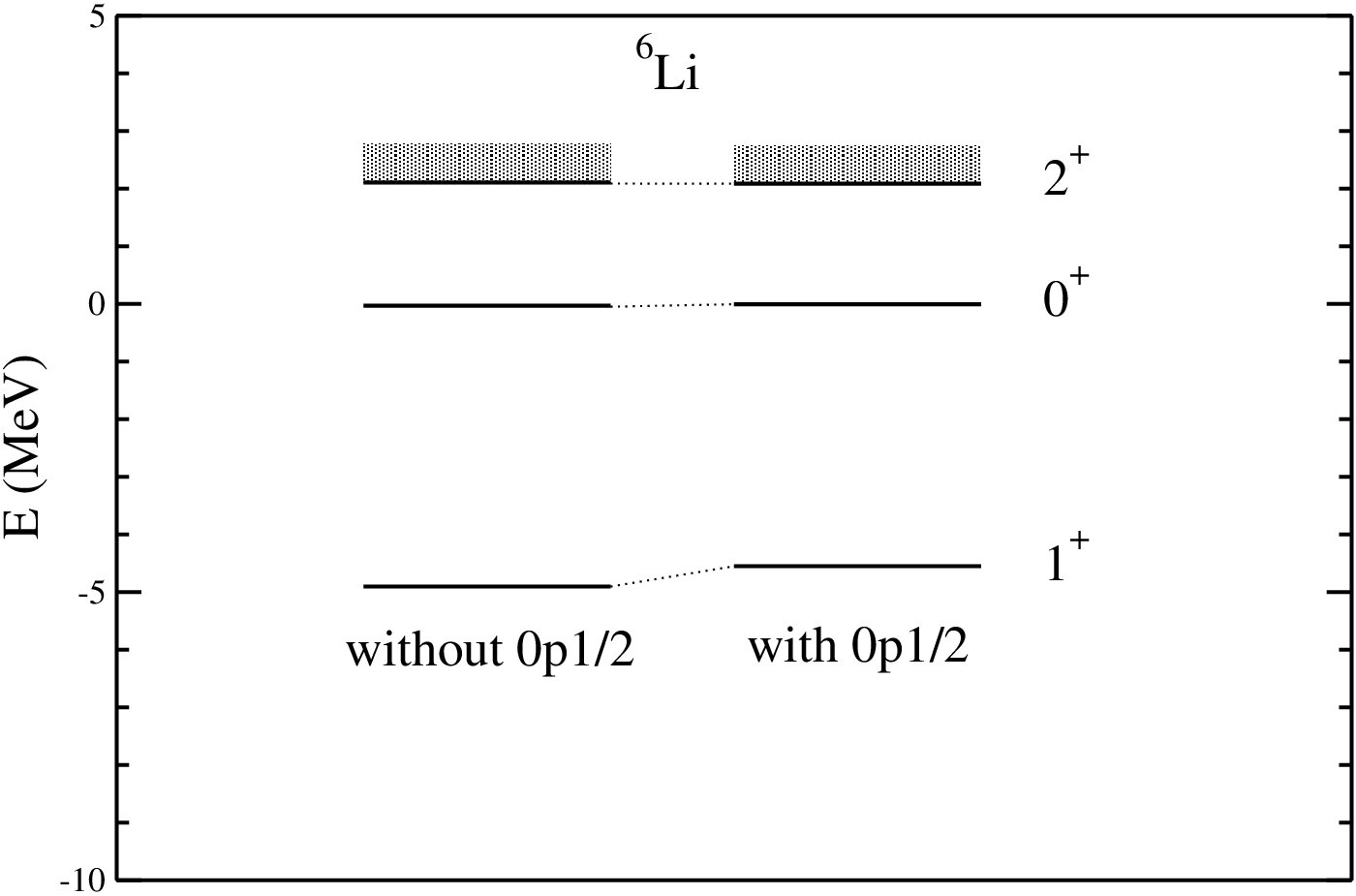}
\end{center}
\caption{Eigenstates of $^6$Li calculated with and without the $0p_{1/2}$
         resonant state in the basis.} 
\label{Li6_comparison}
\end{figure}

To calculate the first $2^+(T=0)$ of $^6$Li and in order to have the two first $5/2^-$ of $^7$Li, nevertheless, 
it is necessary to take into account the $0p_{1/2}$ states, even if they are broad. 
Indeed, without these states, the first $2^+(T=0)$ of $^6$Li does not exist at the level of pole approximation
and as a consequence is impossible to find. Moreover, there is only one $5/2^-$ for $^7$Li at the level of pole approximation
without the $0p_{1/2}$ states.
As a consequence, one used the resonant $0p_{1/2}$ states in the basis to calculate these two states, thus deforming
the $p_{1/2}$ contours in the complex plane so the resonant $0p_{1/2}$ states are enclosed. One also took twice as more points
for these contours in order to further reduce discretization effects.
To check that the two set of basis states are equivalent, one calculated the first $1^+$, $0^+$ and $2^+(T=1)$ with them.
The first $3^+$ is not mentioned as no $p_{1/2}$ state, resonant or scattering, can enter its decomposition for obvious geometrical reasons.
The comparison between the two is shown in Fig. \ref{Li6_comparison}, where it is clear that the two ways to calculate the eigenstates are equivalent.

\subsection {The lithium chain}    
 
As a pilot example of GSM calculations in the space of proton {\it and} 
neutron states, 
we have chosen to investigate  the Li chain. The
continuum effects are very important in these nuclei, 
both in their  ground  states and in excited states. 
The nucleus $^{11}$Li is also a 
well-known example of a two-neutron halo. In our $p$-space(s) calculation,
 we  consider the  one-body Coulomb potential of the $^{4}$He core, 
which is given by a uniformly charged sphere
having  the radius of the WS potential. It turns out that the inclusion of
the one-body Coulomb potential  modifies the GHF basis in
lithium isotopes as compared to the helium isotopes, an effect which is
usually neglected in the standard SM calculations.

Recent studies of the binding energy systematics in the
$sd$-shell nuclei using the
Shell Model Embedded in the Continuum (SMEC) \cite{(Ben99),(Ben00c)} have reported
a significant reduction of the neutron-proton $T$=0
 interaction with respect
to the neutron-neutron $T$=1 interaction  in the nuclei
close to the neutron drip line \cite{(Luo02),(Mic04)}. In SMEC, 
this reduction is associated with  a  decrease in the
one-neutron emission threshold when approaching the neutron drip line, 
i.e.,  it is a genuine continuum coupling effect. The
detailed studies in fluorine isotopes have shown that the  reduction of 
the $T$=0
neutron-proton interaction {\em cannot} be corrected by 
any adjustment of the monopole components  of the effective
Hamiltonian. To account for  this effect  in the standard 
SM, one would need to introduce a particle-number-dependence of the $T$=0 
monopole terms.
Interestingly, it has recently been suggested \cite{(Zuk03)}
that a linear reduction of $T$=0 two-body 
monopole terms  is expected if one incorporates  three-body
interactions into  the two-body framework of a standard SM.

Our  GSM studies of lithium isotopes indicate that the reduction of
$T$=0 neutron-proton interaction with increasing neutron number is
essential. For example, if one uses the $V_0(J,T=0)$ strength adjusted 
to $^{6}$Li to calculate $^{7}$Li, the g.s. of $^{7}$Li becomes overbound by 13
MeV, and the situation becomes even worse for heavier Li isotopes. To reduce this
disastrous tendency, in the first approximation, we have used a linear dependence
of $T$=0 couplings on the number of valence neutrons $N_n$:
\begin{eqnarray}
\label{inter}
V_0(J=1,T=0)=\alpha_{10}\left[1-\beta_{10}(N_n-1)\right],   \\ 
V_0(J=3,T=0)=\alpha_{30}\left[1-\beta_{30}(N_n-1)\right],
\end{eqnarray}
with $\alpha_{10}=-600$ MeV fm$^3$, $\beta_{10}=-50$ MeV fm$^3$,
$\alpha_{30}=-625$ MeV fm$^3$, and $\beta_{30}=-100$ MeV fm$^3$. 
This linear dependence is probably oversimplified, as shown in
Refs.~\cite{(Luo02),(Mic04)} where the proton-neutron $T$=0 interaction first
decreases fast with increasing neutron number and then saturates for weakly
bound systems near the neutron drip line. For the $T$=1
interaction, we have taken parameters $V_0(J=0,T=1)$ and 
$V_0(J=2,T=1)$ determined for the  He ground states \cite{(Mic04a)}, as they provide reasonable results
in the He chain.

\begin{figure}[htbp]
\includegraphics[width=15cm]{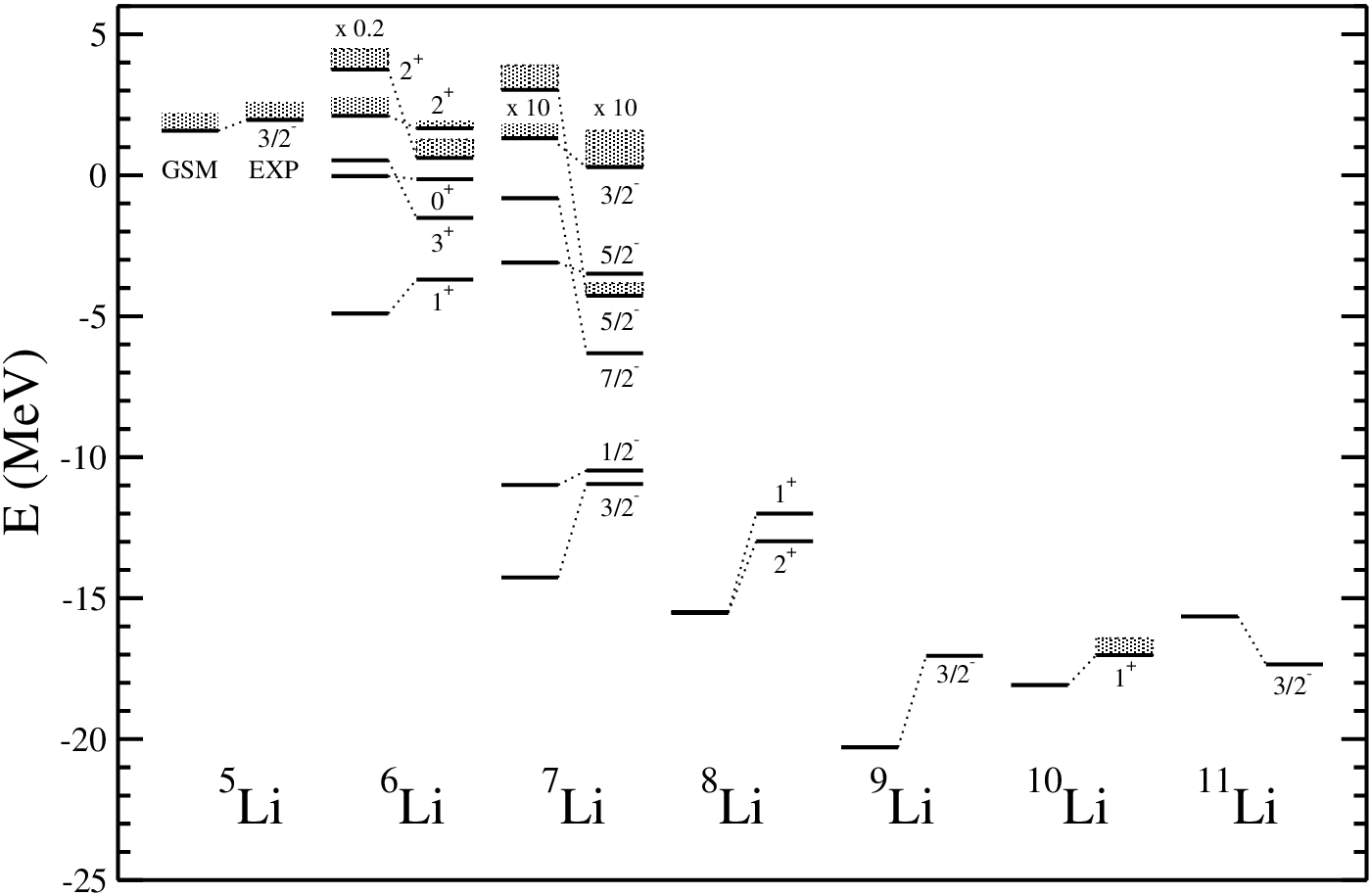}
\caption{Experimental (EXP) and predicted (GSM)  binding energies and 
spectra of lithium isotopes obtained with the SGI Hamiltonian.
The resonance widths are indicated by shading.
The energies are given with respect to the core of $^{4}$He.
Experimental data are taken from \cite{(Boh97),(ensdf),(masses)}.}
\label{Li_spectrum}
\end{figure}
The results of our GSM calculations for the neutron-rich Li isotopes are
shown in Fig.~\ref{Li_spectrum}. One obtains a reasonable description of
the g.s. energies of lithium isotopes relative to the g.s. energy of
$^{4}$He, but excited states are  reproduced roughly. Clearly, the
particle-number dependence of the matrix elements has to be further
investigated in order to achieve the detailed description of the data.
The absence of an anti-bound $s_{1/2}$ state in the Berggren basis is
also likely responsible for large deviations with the data seen for 
$^{10}$Li and $^{11}$Li. 

In a number of cases, excited GSM
states are calculated to lie above several
decay thresholds, i.e., they are predicted to be unstable to,
single-nucleon, deutron, proton+neutron emission, and/or $\alpha$ decay.
The total decay width of a nucleus in a given GSM eigenstate is given by
an imaginary part of its complex eigenenergy. Different open decay
channels contribute incoherently to the total decay width and their
respective partial width cannot be separated easily
\cite{(Oko03)}. In practical
applications, however, one may calculate  the spectroscopic factors for
the separation of nucleon(s) or nucleon groups in a given GSM
eigenstate, following the well-known procedures of the standard
shell model \cite{(Bal59),(Wil62)}, i.e., by calculating the probability to find a certain
one- or many-particle configuration formed by $A-k$ and $k$ nucleons in a
state  of the $A$-nucleon system. We intend to implement this
option in the future.

\begin{table}
\caption{\label{isospin_table}Average isospin (\ref{isospav})  
calculated in GSM for various states 
in $^{6,7}$Li.}
\parbox{13cm}{
\begin{ruledtabular}
\begin{tabular}{|c|cccccccc|} 
$J^\pi$  & 1$^+$ & 2$_1^+$        & 2$_2^+$    & 3/2$^-$ (g.s.) & 1/2$^-$ & 5/2$^-$ & 7/2$^-$ & 3/2$^-$ \\ \hline
$T_{av}$ & 0.01  & 0.023+i0.020   & 1+i0.015   & 0.509          & 0.514   & 0.507   & 0.505   & 1.503+i0.011
\end{tabular}
\end{ruledtabular}
}
\end{table}
Due to the explicit  presence
of the  Coulomb potential
in the GSM,  isospin  is no longer conserved. In order to assess the isospin-mixing
effect, for each shell model state $| \Psi \rangle$ we define  the average isospin quantum number 
$T_{av}$ in the following way:
\begin{equation}\label{isospav}
T_{av} = \frac{-1 + \sqrt {1 + 4 \langle \Psi | \hat{T}^2 | \Psi \rangle} }{2}
\end{equation}
where the  isospin  raising and lowering operators  in $\hat{T}^2$
act on 
single-proton and single-neutron states with explicitly {\em different} asymptotic behavior.
As seen in Table~\ref{isospin_table},
the  values of $T_{av}$ indicate  very small isospin-mixing effects. 
The imaginary part for  2$^+$ of $^6$Li and 3/2$^-$ of $^7$Li 
comes from the
fact these states are unbound.
This result demonstrates that despite the presence of the Coulomb potential
 and despite a large coupling to the
continuum in some cases, the isospin quantum number is still almost nearly
conserved. That is, isospin is a very good characteristic
 of  nuclear states, even if they are unbound, such as
 the second $3/2^{-}$ state 
of $^7$Li (which is a $T$=3/2 isobaric analogue  of the $^7$He ground
state) or the lowest $T$=1, $J^\pi$=2$^+$ state of $^6$Li
(which can be viewed as a  $T$=1 isobaric analogue 
of the first excited state of $^6$He).

The identification of states can be problematic when the eigenstates have same angular momenta and parities.
When their isospin is different, the calculation of the approximate isospin quantum number is useful to identify
one state with its experimental counterpart, as was done with the two $2^+$ states of $^6$Li.
For the two $5/2^-$ of $^7$Li, which have both $T=1/2$ experimentally, the consideration of their
width can be used instead to identify them. Indeed, one of them is broad with a width of 880 keV, while
the other is narrower with a width of 89 keV. As one $5/2^-$ state has to occupy $0p_{1/2}$ broad resonant
states at the level of pole approximation while the other does not, the latter state can be associated
with the $5/2^-$ with a width of 89 keV, while the former can be associated with the $5/2^-$ with a width of 880 keV.

\section{Conclusions}\label{concludo}
The Gamow Shell Model, which has been 
introduced only very recently \cite{(Mic02),(Mic03)}, has proven to be a reliable tool
for the microscopic description of weakly bound and unbound nuclear states. In the He 
isotopes, GSM, with either SDI or SGI interactions, was able to describe 
fairly well binding energy patterns and low-energy spectroscopy,
in particular the Borromean features in
the chain $^{4-8}$He \cite{(Mic03),(Mic04a)}. Using the finite-range SGI interaction made it 
possible to perform  calculations in the GHF basis, thus  designing the optimal
Berggren basis for each nucleus. 
 
In the Li isotopes, the results crucially depend on the
$T$=0 interaction channel. It was found that the $T$=0 force
should  contain a pronounced density (particle-number)
dependence which originates from the
coupling to the continuum and leads to an effective renormalization of the
neutron-proton coupling. This effect cannot be absorbed 
by the modification of $T$=0 monopole terms in the standard SM
framework. The effective 
renormalization of ($J=1,T=0$) and ($J=3,T=0$) couplings, and, to a lesser extent, 
other coupling constants
found in
the present GSM studies has to be further investigated.
To better take into account all these effects, calculations with a finite-range,
density-dependent   interaction
inspired by the Gogny force \cite{(Dec80)} are now in progress. 
The three main problems related to the realistic GSM 
calculations are  the treatment
of the center of mass (essential, especially in the context of halo nuclei),
the inclusion of anti-bound states \cite{(Tho94),(Bet04),(Ver89),(Hag02)},
and the handling of very large shell-model spaces.

For that matter, the successful application of GSM  to heavier nuclei is ultimately related to
the progress in  optimization of the GSM basis, related to 
the inclusion of the non-resonant continuum configurations. A promising 
development is the adaptation of
the density matrix renormalization group  method \cite{(Whi93)} to the genuinely
non-hermitian SM problem in the complex-$k$ plane
using the $j$-scheme  \cite{(Rot04),(Mic04a)}.

In summary, in this paper we report the  
proof-of-principle proton-neutron calculations
using the Gamow Shell Model. Our single-particle proton Hamiltonian contains 
the Coulomb term that explicitly breaks isospin symmetry.
 In order to extend GSM calculations
to open-proton systems, the Berggren completeness relation has been extended to
the case of the Coulomb potential.
The completeness relation has also been derived for nonlocal interactions
that naturally appear in the  GHF method, a Hartree-Fock
inspired procedure to optimize the s.p. basis.
 According
to GSM, the  isospin mixing-effects are very weak, even for high-lying unbound states of
Li isotopes.

\begin{acknowledgments}
Discussions with Akram Mukhamedzhanov  are gratefully acknowledged.
This work was supported in part by the U.S. Department of Energy
under Contracts Nos.  DE-FG02-96ER40963 (University of Tennessee),
DE-AC05-00OR22725 with UT-Battelle, LLC (Oak Ridge National
Laboratory), and DE-FG05-87ER40361 (Joint Institute for Heavy Ion
Research).
\end{acknowledgments}


\end{document}